\documentclass[runningheads]{llncs}
\usepackage{multirow}
\usepackage{graphicx}
\usepackage{amsmath}
\long\def\comment#1{}

\long\def\comment#1{}

\usepackage{xspace}
\newcommand*{\eg}{e.g.\@\xspace}
\newcommand*{\ie}{i.e.\@\xspace}
\newcommand*{\etal}{et al.\@\xspace}

\begin{document}
\title{Dual Adversarial Learning with Attention Mechanism for Fine-grained Medical Image Synthesis}
%
%
\author{Dong Nie$^{1,2}$, Lei Xiang$^3$, Qian Wang$^3$, Dinggang Shen$^1$%
}
\institute{
$^1$Department of Radiology and BRIC, University of North Carolina at Chapel Hill, USA\\
$^2$Department of Computer Science, University of North Carolina at Chapel Hill, USA\\
$^3$Med-X Research Institute, School of Biomedical Engineering, Shanghai Jiao Tong University, Shanghai, China\\
}
%
%
%
\maketitle              
\begin{abstract}
Medical imaging plays a critical role in various clinical applications. However, due to multiple considerations such as cost and risk, the acquisition of certain image modalities could be limited. To address this issue, many cross-modality medical image synthesis methods have been proposed. However, the current methods cannot well model the hard-to-synthesis regions (\eg, tumor or lesion regions). To address this issue, we propose a simple but effective strategy, that is, we propose a dual-discriminator (dual-D) adversarial learning system, in which, a global-D is used to make an overall evaluation for the synthetic image, and a local-D is proposed to densely evaluate the local regions of the synthetic image. More importantly, we build a difficult-region-aware attention mechanism which targets at better modeling hard-to-synthesize regions (\eg, tumor or lesion regions) based on the local-D.
Experimental results show the robustness and accuracy of our method in synthesizing target images from the corresponding source images. In particular, we evaluate our method on two datasets, \ie, to address the tasks of generating T2 MRI from T1 MRI for the brain tumor images and generating MRI from CT. Our method outperforms the state-of-the-art methods under comparison in all datasets and tasks. And the proposed difficult-region-aware attention mechanism is also proved to be able to help generate more realistic images, especially for the hard-to-synthesize regions.

\end{abstract}
\section{Introduction}
\label{sec:intro}
The importance of medical imaging for clinical diagnosis, treatment of disease, and medical
research has steadily risen over the last decades. Multiple imaging modalities provide complementary information for each other, which is essential for comprehensive assessment of complex diseases in either diagnostic examinations or as part of medical research trials.
For example, computed tomography (CT) is often used for dose planning in radiation therapy for cancer patients. However, CT scan will expose patient to radiation and its image cannot provide good contrast in soft tissue. In contrary, magnetic resonance imaging (MRI) provides very good soft tissue contrast and is much safer than CT (as no radiation is involved during acquisition). However, MRI is not directly related to tissue density information which is often required for radiotherapy planning or PET image reconstruction~\cite{kinahan1998attenuation}. Based on the above observations, we argue that each modality is needed at different stages during disease diagnosis and treatment, while it is not easy to obtain them all in practice. To this end, it can be very beneficial to study the solutions that synthesizing modality of interest (or target) from the available source modalities. It provides the modality data required for many clinical trials without additional cost/risk of performing the actual acquisition.

\begin{figure*}[t!]
\centering
  \includegraphics[width=1\linewidth]{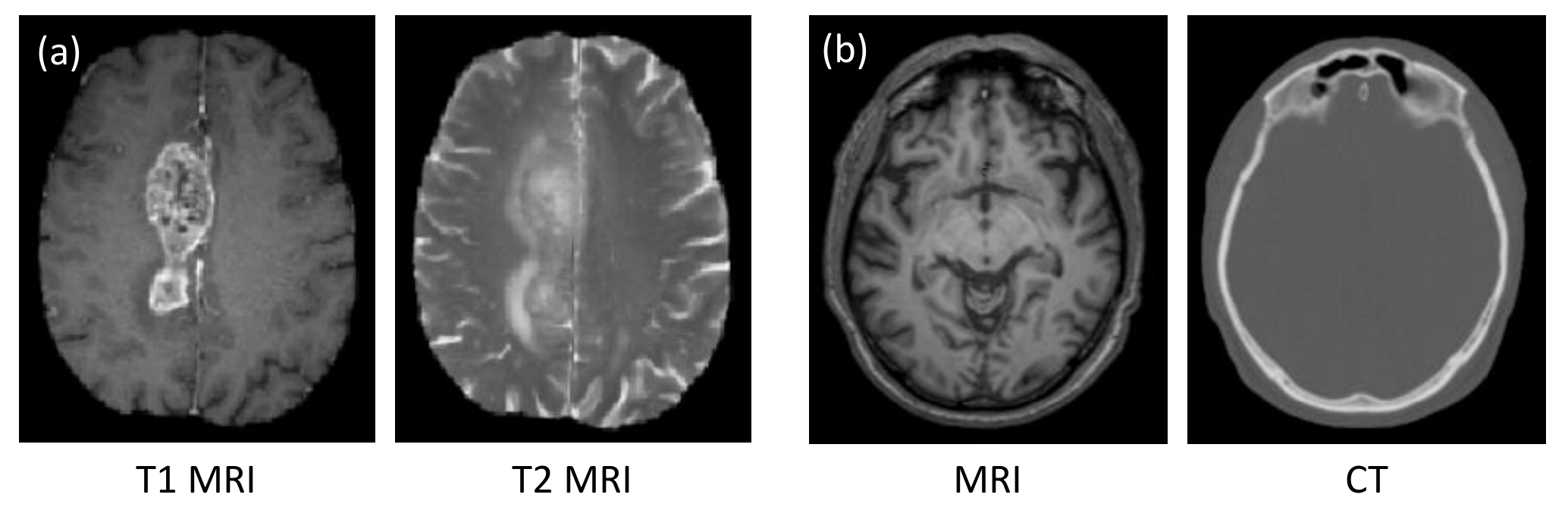}\\
  \caption{Three pairs of corresponding source (left) and target (right) images from the same subjects. (a) shows a pair of T1 MRI/T2 MRI brain tumor images; (b) shows a pair of MRI/CT brain images.}
\label{fig:example}
\end{figure*}

Many researchers have tried to directly synthesize high-quality demanded medical modality images~\cite{han2017mr,nie2017}. However, it is quite challenging due to the following possible problems. Firstly, the mapping from the source modality to the target modality (or its inverse) is typically complex and ill-posed~\cite{hefnawy_sr_challenges}. Moreover, different modalities may show quite different image appearances, \eg, MRI and CT as shown in Fig.~\ref{fig:example}(b). Furthermore, certain regions in the image have totally different image contrast and even different shapes, for instance, the tumor regions in Fig.~\ref{fig:example}(a). Despite all these challenges, there are potential connections between the two modalities if we observe deep enough. That is, the mapping from source modality to target modality should be highly non-linear so that it can bridge the significant appearance gap between the two modalities.

Convolutional neural network (CNN) provides a natural way for learning highly non-linear mapping because of the multiple layer settings. In the particular case of image synthesis, Dong \etal~\cite{dong2014learning} proposed to use CNNs for single image super-resolution. Li~\etal~\cite{li2014deep} applied a similar deep learning model to estimate the missing PET image from the MRI data of the same subject. Huang~\etal~\cite{huang2017simultaneous} proposed to simultaneously conduct super-resolution and cross-modality medical image synthesis by the weakly-supervised joint convolutional sparse coding. Nie~\etal~\cite{nie2017} proposed supervised adversarial learning framework with gradient difference loss to synthesize CT from MRI.

Although the training of the above mentioned image synthesis methods could achieve good performances in most cases, they cannot obtain reasonable results in certain situations, such as tumor regions when dealing with disease diagnosis shown in Fig.~\ref{fig:example}(a), since it does not consider samples or regions that are difficult though very important to process. The deep reason behind is that the training of the network tends to be dominated by samples/regions that are easy to synthesize, \ie, normal tissue regions. This easy-to-synthesize sample/region dominance phenomenon often occurs in medical image synthesis tasks due to the irregular tumor/lesion distribution. These tumor/lesion regions may be ignored owing to the relatively small size in the whole image, although they are the most important biomarkers for clinical diagnosis. Thus, it is quite important to develop a method that could better model the lesion/tumor regions in medical image synthesis tasks.

In this work, we propose a dual-discriminator adversarial learning framwork with difficult-region-aware mechanism to solve the above mentioned issues. Specifically, besides the regular CNN-based discriminator, we also propose a dense FCN as the local-discriminator to model the easy-or-hard extent for the synthesis task. More importantly, we further propose a difficult-region-aware mechanism to better model the hard-to-synthesis regions (\ie, tumor regions). Experimental results demonstrate that the proposed method can effectively synthesize target images with much better modeling capacity on the hard-to-synthesize regions. To the best of our knowledge, this is the first work that tries to address the hard-to-synthesis regions on cross-modality image synthesis tasks.

\section{Methods}
\label{sec:method}

To address the above mentioned problems and challenges, we propose a deep convolutional adversarial network framework by adversarially training the generator (UNet) via dual discriminators: a CNN as the global discriminator and an FCN as the local discriminator, the framework of which is shown in Fig.~\ref{fig:gan}. First, we propose a basic UNet generator to estimate the target image from the corresponding source image. Note that we adopt 2.5-D operations to better model the spatial mapping and thus could alleviate the discontinuity problem across 2D slices. Second, we utilize the adversarial learning strategy~\cite{goodfellow2014generative} for the designed image synthesis network, where two additional discriminator networks are modeled. The first discriminator (global one) urges the generator`s output to be similar with the ground-truth target image perceptually. The second discriminator (local one) compares the local regions of the synthetic target images to those of the real target images, so that we can obtain local similarities between the synthetic and real images and thus further improve the hard-to-predict regions.
At the testing stage, an input source image is first partitioned into overlapping patches, and, for each patch, the corresponding target is estimated by the generator. Then, all generated target patches are merged into a single image to complete the source-to-target synthesis by averaging over overlapping target regions. In the following, we will describe in detail the proposed medical image synthesis framework.

\subsection{Supervised Generative Adversarial Network}
\label{subsec:gan}

\begin{figure*}[t!]
  \includegraphics[width=1\linewidth]{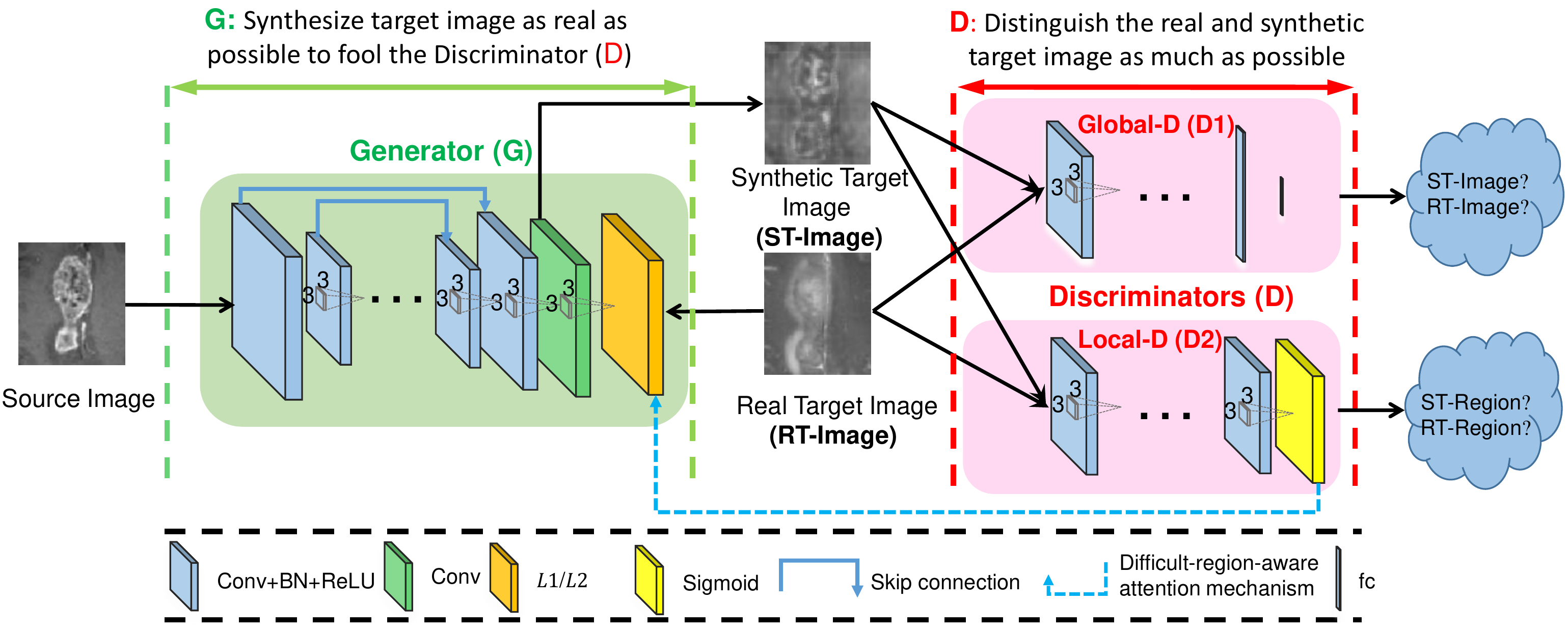}
  \caption{Architecture used in the deep supervised generative adversarial setting to synthesize target image. This framework contains one generator and two discriminators. A difficult-region-aware attention mechanism is also included in the framework.}
  \label{fig:gan}
\end{figure*}

As mentioned above, we propose a supervised deep convolutional adversarial framework, which is inspired by the recent popular generative adversarial networks (GANs)~\cite{goodfellow2014generative}, to complete the source-to-target synthesis as shown in Fig.~\ref{fig:gan}. The components in Fig.~\ref{fig:gan} will be introduced in the following paragraphs.

\subsection{UNet for Medical Image Synthesis}
UNet, an evolutionary version of fully convolutional networks (FCN) which incorporates the high-resolution (low layers) feature maps and highly-semantic (high layers) feature maps to increase the localization accuracy, is widely used for segmentation and reconstruction in both computer vision and medical image analysis~\cite{ronneberger2015u,dong2016image,han2017mr,nie2016estimating,nie2018strainet}. It can preserve spatial information in local neighborhood of the image space and is also much faster compared to CNN at the testing stage. In this paper, we adopt UNet to implement the image generator to perform the medical image synthesis task since UNet would potentially alleviate the loss of resolution compared to FCN.

As mentioned in the Introduction section, typically an $L_1/L_2$ loss is used to train the model as shown in Eq.~\ref{eq:loss_base}.
\begin{equation}
L_G(X,Y)=\left \| Y-G(X) \right \|^p
\label{eq:loss_base}
\end{equation}
where $Y$ is the ground-truth target image, and $G(X)$ is the generated target image from the source image $X$ by the Generator network $G$ and $p$ is $1$ or $2$.

\subsection{Adversarial Learning for Medical Image Synthesis}
To make the generated target images better perceptually~\cite{goodfellow2014generative,radford2015unsupervised,nie2017}, we propose to use adversarial learning to improve the performance of UNet. Instead of using classical CNN-based discriminator to globally evaluate how well the images are generated, we propose using an additional dense FCN-based local discriminator to capture the quality of the synthetic images for local regions.

Specifically, our networks include 1) the generator for estimating the target image and 2) the global discriminator for distinguishing the real target image from the generated one and a local discriminator distinguish synthetic and real images in voxel-level, as shown in Fig.~\ref{fig:gan}. The generator network $G$ is an FCN as described above. The global discriminator network $D1$ is a CNN, following the popular Wasserstein GAN ~\cite{arjovsky2017wasserstein,gulrajani2017improved}. The local discriminator is a FCN with a dense output, each element of which corresponds to the probability of how well the local region around this voxel is synthesized. It is north noting that global-D and local-D are independent. $D1$ and $D2$ are trained simultaneously, while $G$ are iteratively updated with these two discriminators.

\noindent{\bf Global Adversarial Learning: }
The classical GAN~\cite{goodfellow2014generative} may lead to training failure due to the limitation of the Jensen–Shannon divergence (JS-divergence)~\cite{arjovsky2017wasserstein}.
To address this issue, WGAN~\cite{arjovsky2017wasserstein} uses the Wasserstein distance instead of the JS-divergence to compare
data distributions. Later, gradient penalty strategy(WGAN-GP)~\cite{gulrajani2017improved} is further proposed to replace the gradient clipping in the WGAN.

In our study, we follow the WGAN-GP to form the global discriminator ($D1$). Concretely, the loss function for $D1$ can be defined as:

\begin{equation}
L_{D1}(X,Y) = E_x[D1(G(X))]-E_Y[D1(Y)]+\lambda E_{\hat{X}} [(\left \| \bigtriangledown_{\hat{X}} D1(\hat{X})\right\|_2-1)^{2}]
\label{eq:loss_D1}
\end{equation}
where $X$ is the source input image, $Y$ is the corresponding target image, $G(X)$ is the estimated image by the generator, and $\hat{X}$ is uniformly sampled along straight
lines among synthetic and real samples; and
$\lambda$ is a constant weighting hyper-parameter.

On the other hand, the global adversarial loss term used to train $G$ is defined as Eq.~\ref{eq:adv1}.
\begin{equation}
\label{eq:adv1}
L_{G\_ADV1}(X,Y) = -E_x[D1(G(X))]
\end{equation}

 With the above equations, $D1$ can globally distinguish the real target data and the synthetic target data synthesized by $G$. At the same time, $G$ aims to produce more realistic target images and to confuse $D1$. The architecture of $D1$ follows the suggestions in~\cite{gulrajani2017improved,arjovsky2017wasserstein}.

\noindent{\bf Local Adversarial Learning:}
The training objective of the confidence network is the summation of binary cross-entropy loss over the image domain, as shown in Eq.~\ref{eq:loss_D}. Here, we use $G$ and $D2$ to denote the generator and local-D networks, respectively.

\begin{equation}
L_{D2}(\mathbf{X},\mathbf{Y};\mathbf{{\theta _{D2}}})=L_{BCE}(D2(\mathbf{Y},\mathbf{{\theta _{D2}}}),\mathbf{1})+L_{BCE}(D2(G(\mathbf{X}),\mathbf{{\theta _{D2}}}),\mathbf{0}),
\label{eq:loss_D}
\end{equation}
where
\begin{equation}
\begin{aligned}
{L_{BCE}}\left( {\mathbf{\widehat Q},\mathbf{Q}} \right) =  - \sum\limits_{h = 1}^H {\sum\limits_{w = 1}^W {{Q_{h,w}}\log \left( {{{\widehat Q}_{h,w}}} \right)} } + \left( {1 - {Q_{h,w}}} \right)\log \left( {1 - {{\widehat Q}_{h,w}}} \right)
\label{eq:bce}
\end{aligned}
\end{equation}
where \({\mathbf{X}}\) and \({\mathbf{Y}}\) represent the input data and its corresponding real target image, respectively. \({\mathbf{\theta _{D2}}}\) is network parameters for the local-D network.

For training the generator network, besides the $L_1/L_2$ loss defined in Eq.~\ref{eq:loss_base} and the global adversarial learning loss in Eq.~\ref{eq:loss_adv2}, the local adversarial loss (``ADV'') to improve $G$ and fool $D2$ can be defined by Eq.~\ref{eq:loss_adv2}:
\begin{equation}
{L_{ADV2}}\left( {\mathbf{X},\mathbf{{\theta _G}}} \right) = {L_{BCE}}\left( {D2\left( {G\left( {\mathbf{X};\mathbf{{\theta _G}}} \right)} \right),\mathbf{1}} \right)
\label{eq:loss_adv2}
\end{equation}

The training of the two networks is performed in an alternating fashion. First, $D$ is updated by taking a mini-batch of real target data and a mini-batch of generated target data (corresponding to the output of $G$). Then, $G$ is updated by using another mini-batch of samples including sources and their corresponding ground-truth target images.

\subsection{Region-attention based Adversarial Difficulty Learning}
\label{subsec:ssl}
\vspace{-5pt}
Due to the inhomogeneous characteristics and irregular distribution of the medical images, certain region of the images are usually more difficult to well synthesize. As a consequence, it is quite desired to build a model that can better model the hard-to-prediction regions. Since the local discriminator could provide the local confidence information about how well each region is synthesized, we can thus pay more attention on the hard-to-predict regions (\eg, lesion regions) so that these regions can be better modeled. To this end, we propose a adversarial difficulty-aware attention mechanism to better represent the easy-or-hard information. Specifically, we design a difficulty-aware $L_1/L_2$ loss using region-level attentions from the adversarial local confidence map.


The voxel-level difficulty-aware attention from the confidence map ($M$) is formulated (based on Eq.~\ref{eq:loss_base}) in Eq.~\ref{eq:wloss_base}:
\begin{equation}
L_{AttG}(X,Y)=F \odot \left \| Y-G(X) \right \|^p
\label{eq:wloss_base}
\end{equation}
where $\odot$ is the element-wise multiplication and
\begin{equation}
F = {\left( {1 - M} \right)^\beta }
\end{equation}
where \(\beta \) is the voxel-level attention parameter. Note, $F$ here works as a scaling factor, which largely suppresses the contribution of easy-to-synthesize regions to the training loss and emphasize the hard-to-synthesize regions.

With the difficult-region-aware $L_1/L_2$ loss in Eq.~\ref{eq:wloss_base}, we can pay more attention in the less confidently (\ie, hard-to-predict) regions and thus better model them (\eg, tumor or lesion regions). And this adversarial difficulty-region-aware attention mechanism provides an opportunity to use voxel-wise focal loss in regression context.

\noindent{\bf Total Loss for Training Generator:} To this end, the total loss for training generator includes the attention based $L_1/L_2$ loss, the global adversarial loss, and the local adversarial loss, which can be expressed as Eq.~\ref{eq:totalLoss}.
\begin{equation}
\label{eq:totalLoss}
L_G = L_{AttG} + \lambda_1 L_{G\_ADV1} +  \lambda_2 L_{G\_ADV2}
\end{equation}
The above training loss could encourage $G$ to generate target images following several constraints, \ie, besides the voxel-wise correspondence, it also needs to fool the discriminators both globally and locally.



\subsection{Training Details}
Shown in Fig.~\ref{fig:gan}, this generator takes a source image as the input, and tries to generate the corresponding target image with a typical UNet. The discriminator $D1$ is a typical CNN including three stages of convolution, BN, ReLU and max pooling, followed by one convolutional layer and three fully connected layers where the first two use ReLU as activation functions. The filter size is $3\times 3$, the numbers of the filters are 32, 64, 128, respecitvely, and 256 for the convolutional layers, and the numbers of the output nodes in the fully connected layers are 512, 128 and 1, respectively. The dense discriminator $D2$ is a typical FCN with three down-sampling and three up-sampling stages, followed by a convolutional layer with sigmoid activation. In both of the discriminators, we apply the spectral normalization~\cite{miyato2018spectral} for all the layers except the last one.

We have randomly extracted source domain patches of size $5\times 144 \times 144$, along with their corresponding target domain patches of size $1\times 144 \times 144$, using the same center points, as the paired training samples. All networks were trained using the Adam optimizer, the initial learning rate for $G$ is set as $5\times10^{-3}$, that of $D1$ is set as $1\times10^{-4}$, and that of $D2$ is set as $1\times10^{-3}$. Note that we decrease the learning rate with a rate of $0.5$ for $G$, $0.2$ for both $D1$ and $D2$ every $2$ epochs. The mini-batch size is set to be 10. The generator was trained using $\lambda_1=0.05, \lambda_2=0.1$.
The code is implemented using the pytorch library\footnote{https://github.com/pytorch/pytorch}, and it will be publicly released upon acceptance.

\section{Experiments and Results}
We choose the BRATS dataset to evaluate our proposed method, which is a publicly available dataset of MRI from brain tumor patients~\cite{menze2015multimodal}. A total of 354 pairs of T1 MRI and T2 MRI were assembled, where 200 subjects were used for training and 60 subjects were used for validation, and the rest 94 subjects were reserved for testing (Note, the dataset is randomly partitioned). The BRATS dataset is acquired under different scanning protocols on separate sites. Thus, the image sizes and resolutions are different across subjects.

To demonstrate the advantage of the proposed method in terms of prediction accuracy, we compare it with three widely-used approaches: atlas-based method~\cite{vercauteren2009diffeomorphic}, FCN~\cite{long2015fully}, UNet~\cite{han2017mr}, and sGAN~\cite{nie2017} (For simplicity and fair comparison, we have removed auto-context refinement and use UNet as the generator). The experiments for the BRATS dataset are only once following the partition mentioned before. And the experiments for the brain MRI-to-CT dataset are done in a leave-one-out fashion. The evaluation metric is the mean absolute error (MAE) and the peak signal-to-noise ratio (PSNR).

\subsection{Impact of Proposed Dual-Discriminator Strategy}
To show the contribution of the proposed dual-discriminator strategy, we conduct comparison experiments between the sGAN with global discriminator (sGAN-1) and sGAN with local discriminator (sGAN-2) and the proposed dual-discriminator strategy (sGAN-dual) on the BRATS dataset. The PSNR values are $27.3$dB, $27.2$dB and $27.6$dB for these three strategies, respectively. Note that these results are achieved with the ordinary $L_1$ loss for the generator. Actually, besides the explanation from local and global adversarial constraints, we can explain it from another view, \ie, the adversarial gradient vanishing issue can be greatly alleviated due to the dual-discriminator system which can thus better guarantee the adversarial learning.

\subsection{Impact of Difficult-Region-Aware Attention Mechanism}
To show the impact of the proposed difficult-region-aware attention mechanism, we first conduct experiments to compare the performance without this mechanism and with this mechanism on the BRATS dataset, and the experimental results indicate that the performance could be improved by $0.2$dB in terms of PSNR using the proposed attention mechanism. To further investigate the impact of the proposed mechanism, we focus on evaluating the synthesis performances only on tumor regions. By using the segmentation ground truth maps on this dataset, we have computed the PSNR on the testing sets only on the tumor regions, which indicates that the PSNR on tumor regions is improved by $0.6$dB in average.

\begin{figure}[t!]
\centering
  \includegraphics[width=1\linewidth]{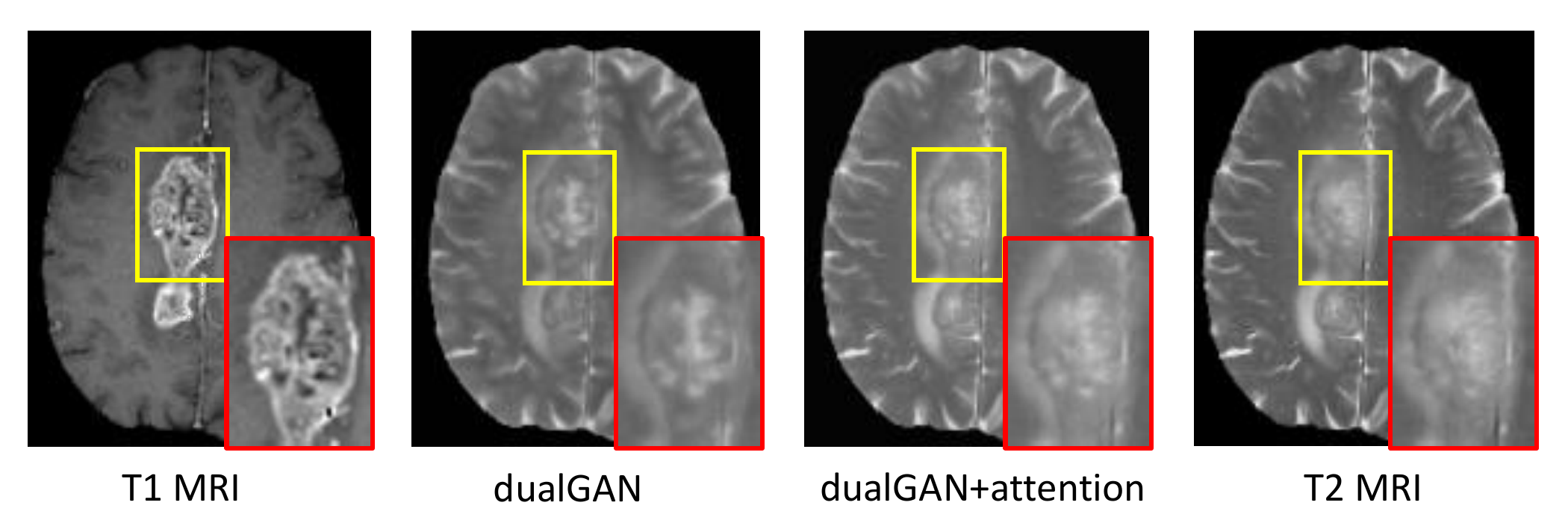}\\
  \caption{Visual comparison for impact of the proposed difficult-region-aware attention mechanism.}
\label{fig:attention-impact}
\end{figure}

We also visualize results in Fig~\ref{fig:attention-impact}, where the leftmost image is the input T1 MRI, and the rightmost image is the ground-truth T2 MRI. We can clearly see that the generated data using the proposed difficult-region-aware attention mechanism (\ie, `dual-D+attention') could recover much more details compared to the one without this mechanism (\ie, `dual-D'), especially for the tumor regions, which further demonstrate the effectiveness of our proposed difficult-region-aware attention mechanism.

To better understand why the difficult-region-aware mechanism works, we also analyze the confidence map generated by the local discriminator (\ie, $D2$). We find that the tumor regions are evaluated to be poorly synthesized, thus, more attention will be propagated to the tumor regions in the generator network, as a consequence, these regions are then better modeled during training.

\subsection{Comparing with Other Methods}
\begin{figure}[h!]
\centering
  \includegraphics[width=1\linewidth]{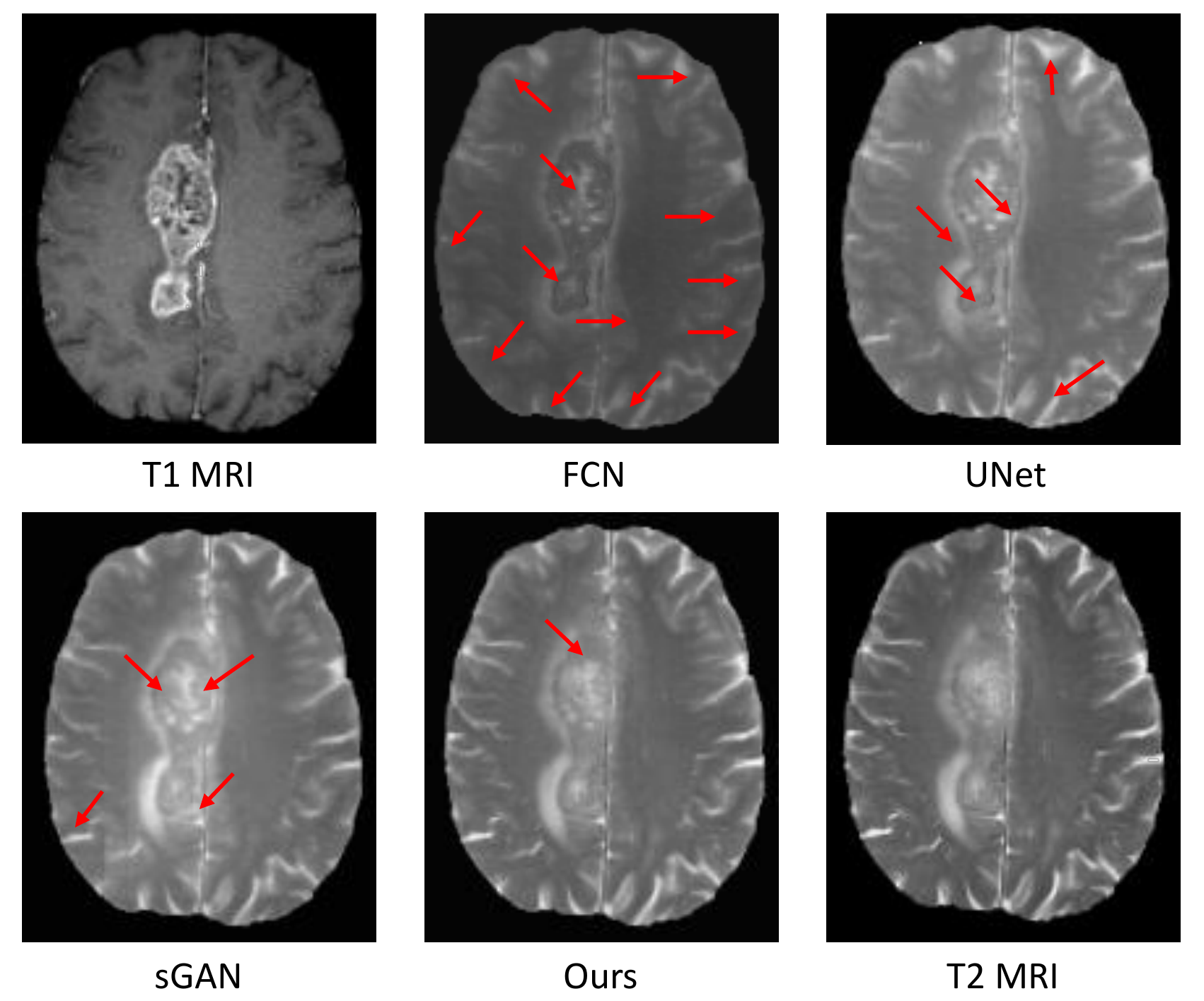}\\
  \caption{Visual comparison of MR image, the estimated CT images by our method and other competing methods, and the ground-truth CT image for the typical brain tumor cases. Red arrows mean poorly synthesized regions.}
\label{fig:brats-res}
\end{figure}

To qualitatively compare the synthetic target image by different methods, we visualize the generated target image with the ground-truth target image in Fig.~\ref{fig:brats-res}. We can see that the proposed algorithm can better preserve the continuity, coalition and smoothness in the synthetic results, since it uses both global and local adversarial learning constraints in the image patch as discussed in Section ~\ref{subsec:gan}. More importantly, the tumor region of generated T1 MRI can recover much more details than other methods, and thus looks much closer to the real T2 MRI compared to all other methods. We argue that this is due to the difficult-region-aware attention mechanism which reweight more on the recognized hard-to-synthesis regions, \ie, tumor regions.

We also quantitatively compare the predicted results in Table~\ref{tab:brats} in terms of PSNR and MAE. Our proposed method outperforms other methods in both metrics, which is consistent with the visual comparison and further demonstrates the advantage of our framework.
\begin{table}[h!]
\begin{minipage}{.5\linewidth}
\centering
\caption{Average MAE and PSNR on 94 \protect\\testing subjects from the BRATS dataset.}
\begin{tabular}{l cc}
\hline
Method & MAE & PSNR\\
\hline
FCN &121.3(18.6)  & 22.7(2.3)  \\
UNet & 107.6(\textbf{14.1})  & 26.0(\textbf{1.4}) \\
sGAN & 100.3(14.7)   & 26.6(1.5)  \\
Ours & {\textbf{92.4}(14.2)}  & {\bf 27.3(1.4)}  \\
\hline
\label{tab:brats}
\end{tabular}
\end{minipage}\begin{minipage}{.5\linewidth}
\centering
\caption{Average MAE and PSNR on \protect\\16 subjects from the brain dataset.}
\begin{tabular}{l cc}
\hline
Method & MAE & PSNR\\
\hline
FCN & 24.4(15.1) & 22.7(3.2) \\
UNet & 21.8(12.8) & 26.7(2.1) \\
sGAN & 20.4(11.2)   & 27.3(\bf{1.7})  \\
Ours & {\bf 18.6(10.3)}  & {\textbf{28.1}(1.8)}\\
\hline
\label{tab:brain}
\end{tabular}
\end{minipage}
\label{tab:res}
\end{table}

\subsection{Validation on Another Brain Dataset}
\begin{figure}[h!]
\centering
  \includegraphics[width=1\linewidth]{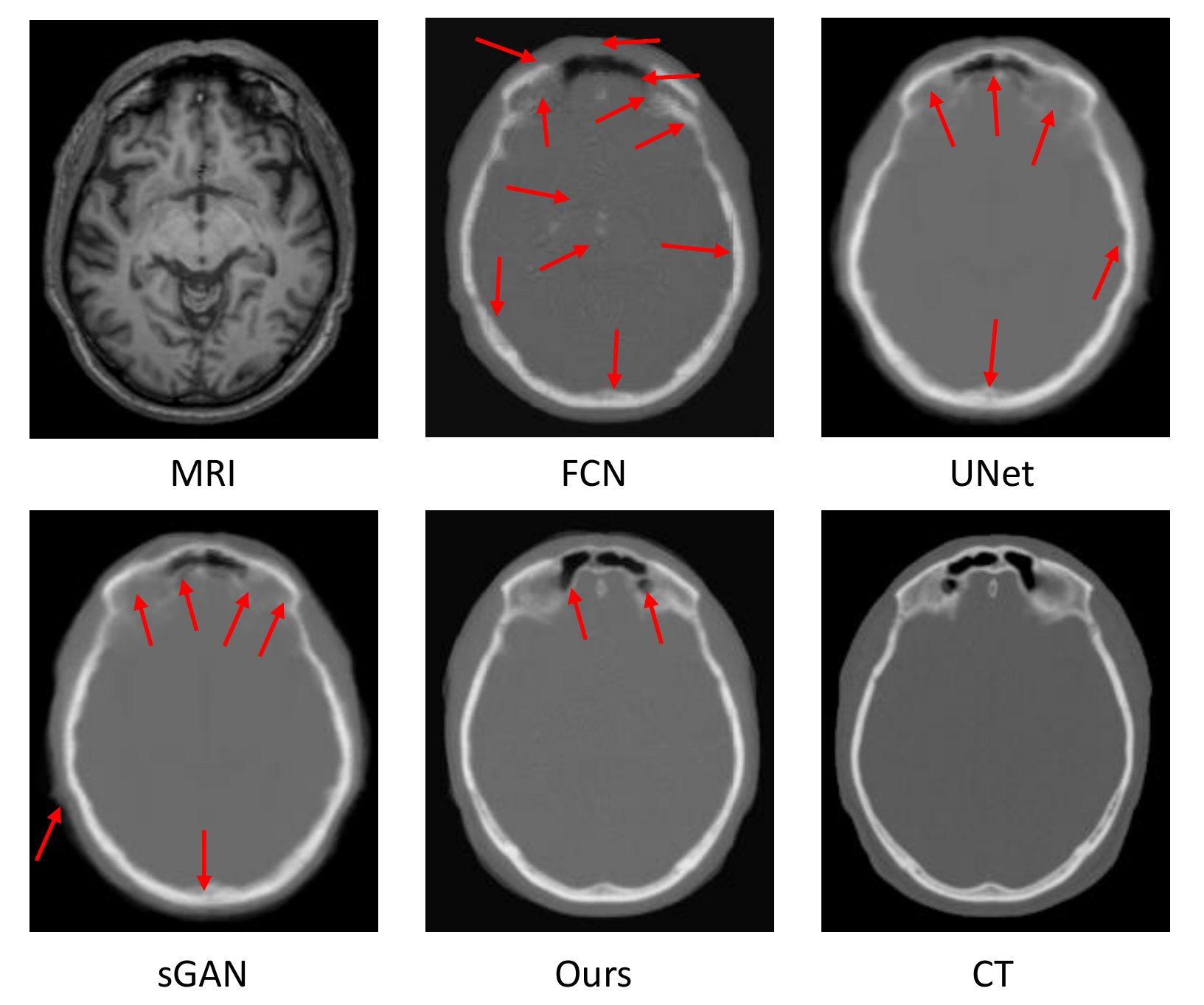}\\
  \caption{Visual comparison of MR image, the estimated CT images by our method and other completing methods, and the ground-truth CT image for the typical brain case. Red arrows mean poorly synthesized regions.}
\label{fig:brain-res}
\vspace{-5pt}
\end{figure}

To show the generalization ability of the proposed method, we also validate the proposed method on the brain dataset, which was acquired from 16 subjects with both MRI and CT scans in the Alzheimer's Disease
Neuroimaging Initiative (ADNI) database (see
\url{www.adni-info.org} for details).
A typical example of preprocessed CT and MR images is given in Fig.~\ref{fig:example}. The qualitative comparison results with different methods are shown in Fig.~\ref{fig:brain-res}, and the quantitative comparison results are shown in Table.~\ref{tab:brain}. Obviously, our proposed method can work better than the state-of-the-art methods, which demonstrate that our proposed method can be generalized to more datasets. More importantly, the proposed method could model much better details after clearly observing the visual results.

\section{Conclusions}
\label{sec:conclusion}
We developed a dual discriminator based adversarial learning framework, including a global one modeling the overall evaluation and a local one modeling the region-wise evaluation, to solve several medical image synthesis tasks. Moreover, we proposed a difficult-region-aware attention mechanism to better model the hard-to-predict regions (\eg, tumor regions). We have applied our proposed model on two tasks, \ie, to predict T2 MRI from their corresponding T1 MRI and to predict brain CT images from their corresponding MR images. The experimental results demonstrate that our proposed method could outperform three state-of-the-art methods. Furthermore, the proposed difficult-region-aware attention mechanism could indeed help improve the hard-to-synthesis regions. Lastly, the proposed method could also be applied to other related medical image synthesis tasks.


\end{document}